\documentclass[preprint,12pt]{elsarticle}
\usepackage{graphicx}
\usepackage[figuresright]{rotating}
\def\tmod{|\tau|}
\newcommand{\be}{\begin{equation}}
\newcommand{\ee}{\end{equation}}
\newcommand{\bey}{\begin{eqnarray}}
\newcommand{\eey}{\end{eqnarray}}

\usepackage{amssymb}

\journal{Computer Physics Communications}

\begin{document}

\begin{frontmatter}

\title{Universal ratios of critical amplitudes in the Potts model
universality class}

\author[lpm]{B. Berche}
\author[infn]{P. Butera}
\author[itpl]{W. Janke}
\author[itp,email]{L. Shchur}

\address[lpm]{Laboratoire de Physique des Mat\'eriaux, Universit\'e Henri Poincar\'e
Nancy 1, BP 239, F-54506 Vand\oe uvre les Nancy Cedex, France}
\address[infn]{Istituto Nazionale di Fisica Nucleare, Sezione di Milano-Bicocca,
Piazza delle Scienze 3, 20126, Milano,   Italia }
\address[itpl]{Institut f\"ur Theoretische Physik, Universit\"at Leipzig,
Postfach 100\,920, 04009 Leipzig, Germany}
\address[itp]{Landau Institute for Theoretical Physics, 142432 Chernogolovka, Russia}
\address[email]{lev@landau.ac.ru}

\begin{abstract}

Monte Carlo (MC) simulations and series expansions (SE) data for the
energy, specific heat, magnetization, and susceptibility of the
three-state and four-state Potts model and Baxter-Wu model  on the
square lattice are analyzed in the vicinity of the critical point in
order to estimate universal combinations of critical amplitudes. We
also form effective ratios of the observables close to the critical
point and analyze how they approach the universal critical-amplitude
ratios. In particular, using the duality relation, we show
analytically that for the Potts model with a number of states $q\le
4$, the effective ratio of the energy critical amplitudes always
approaches unity linearly with respect to the reduced temperature.
This fact leads to the prediction of relations among the amplitudes
of correction-to-scaling terms of the specific heat in the low- and
high-temperature phases. It is a common belief that the four-state
Potts and the Baxter-Wu model belong to the same universality class.
At the same time, the critical behavior of the four-state Potts
model is modified by logarithmic corrections while that of the
Baxter-Wu model is not. Numerical analysis shows that critical
amplitude ratios are very close for both models and, therefore,
gives support to the hypothesis that the critical behavior of both
systems is described by the same renormalization group fixed point.
\end{abstract}

\begin{keyword}
Potts model; Baxter-Wu model; Critical exponents; Critical
amplitudes; Universality; Monte Carlo simulations; Series
Expansions; Renormalization Group
 \PACS 0.50.+q, 75.10.-b
\end{keyword}

\end{frontmatter}


The fixed points of the renormalization group define the universal
behavior of a system through a set of critical exponents and universal
combinations of critical
amplitudes~\cite{PrivmanHohenbergAharony91}. The universality concept
divides all systems at criticality  into a number of universality classes.
It is instructive to know the set of values of the critical exponents and of
the universal combinations of critical amplitudes for a given
universality class.

The two-dimensional Potts model~\cite{Potts52} is the simplest model
which exhibits a phase transition. It is solved exactly at the
critical point for any number of spin components $q$ and it is known
that for $q\le 4$ it has a continuous phase transition while for $q>4$
the phase transition is of the first order. The model has a great
theoretical interest as new theories may be tested in this model.

At the same time, these models may have some practical interest as they
may be realized in an adsorbate lattice placed onto a clean
crystalline surface. The full classification of such systems with
continuous transitions is known theoretically~\cite{Classification}.
There are experiments in which some of them realize the 3-state and
4-state Potts models~\cite{Experiment} and the critical exponents can be
experimentally estimated.

Critical exponents for the Potts model  with $q\le 4$ can be computed exactly
by different theoretical techniques~\cite{Exponents,CFT}. The
values of the thermal critical exponents and of the magnetic
critical exponents follow from the identification of the dimensions of
the conformal algebra operators~\cite{CFT}.

Nowadays, there is no doubt on the values of the leading critical
exponents whereas the values of the correction-to-scaling exponents
are still under discussion, as well as the values of the universal ratios
of the critical amplitudes. Our presentation is intended to give a short
review of the research on the subject.

The Potts model Hamiltonian~\cite{Potts52} (see review~\cite{Wu} for
details) can written as $ H = - \sum_{\langle ij \rangle}\delta_{s_i
s_j}\; ,$ where $s_i$ is a spin variable taking integer values
between $0$ and $q{-}1$, and the sum is restricted to the nearest
neighbor sites $\langle ij \rangle$ on the square lattice.

Close to the critical temperature $T_c$ at which the continuous
phase transition occurs, the residual magnetization $M$ and the
singular part of the reduced susceptibility $\chi$ and of the
 specific heat $C$ of the system in zero external field are
characterized by the critical exponents $\beta$, $\gamma$, and
$\alpha$ and by the critical amplitudes $B$, $\Gamma_\pm$, and
$A_\pm$
\begin{eqnarray}
    M(\tau) &\approx& B (-\tau)^\beta ,\ \tau <0 \label{m-crit}\\
  \chi_\pm(\tau) &\approx& \Gamma_\pm \tmod^{-\gamma}, \label{x-crit}\\
C_\pm(\tau) &\approx& \frac{A_\pm}{\alpha}\tmod^{-\alpha},
    \label{c-crit}
\end{eqnarray}
where $\tau$ is the reduced temperature $\tau=(T-T_c)/T$ and the
labels $\pm$ refer to the high-temperature and low-temperature sides
of the critical temperature $T_c$. The critical amplitudes are not
universal by themselves but some combinations of them, f.e.,
$A_+/A_-$, $\Gamma_+/\Gamma_-$, and $\Gamma_+A_+/B^2$, are
universal~\cite{PrivmanHohenbergAharony91} due to the scaling laws.

On the square lattice, in zero field, the model is self-dual. The
duality relation
\begin{equation}
    \left( e^\beta - 1 \right) \left( e^{\beta^*} - 1 \right)=q
    \label{d-t}
\end{equation}
fixes the inverse critical temperature to $\beta_c{=}\ln
(1{+}\sqrt{q})$. The values $E(\beta)$ and $E(\beta^*)$ of the
internal energy at dual temperatures are simply related through
\begin{equation}
    \left(1-e^{-\beta}\right) E(\beta) + \left(1-e^{-\beta^*}\right)
    E(\beta^*)=2.
    \label{d-et}
\end{equation}

Dual reduced temperatures $\tau$ and $\tau^*$ can be defined by
$\beta{=}\beta_c(1{-}\tau)$ and $\beta^*{=}\beta_c(1{+}\tau^*)$.
Close to the critical point, $\tau$ and $\tau^*$ coincide through
linear order, since $\tau^*{=}\tau{+}\frac{\ln (1{+}\sqrt q)}{\sqrt
q}\tau^2{+}O(\tau^3)$.

The ratio of the free energy critical amplitudes $A_+/A_-$ is equal to
unity due to duality. Moreover, duality relations may be used to
understand the dependence on temperature
of the effective amplitude functions which may
be constructed from the energy in the symmetric phase $E_+(\tau)$
and in the ordered phase $E_-(\tau^*)$
\begin{eqnarray}
    A_+(\tau)&=&\alpha (1-\alpha)\beta_c(E_+(\tau)-E_0)\tau^{\alpha-1}, \label{eff-Aa}\\
    A_-(\tau^*)&=&\alpha (1-\alpha)\beta_c(E_0-E_-(\tau^*))(\tau^*)^{\alpha-1}
    \label{eff-Ab}
\end{eqnarray}
as an effective amplitude ratio
\begin{equation}
\frac{A_+(\tau)}{A_-(\tau^*)}=
\frac{(E_+(\tau)-E_0)\tau^{\alpha-1}}{(E_0-E_-(\tau^*))(\tau^*)^{\alpha-1}},
    \label{rat-e-dual}
\end{equation}
where the constant $E_0$ is the value of the energy at the
transition temperature, $E_0{=}E(\beta_c){=}{-}1{-}1/\sqrt{q}$.

Evaluating expression~(\ref{rat-e-dual}) for small $\tau$ and
denoting $\alpha_q=
-E_0\beta_ce^{-\beta_c}=\frac{\ln(1+\sqrt{q})}{\sqrt{q}}$, we obtain
\begin{eqnarray}
    \frac{A_+(\tau)}{A_-(\tau^*)}&=&1+(3-\alpha)\alpha_q \tau
    +O(\tau^{1+\alpha}).
    \label{d-e-linear}
\end{eqnarray}
\noindent Note the linear dependence on $\tau$ of the effective
amplitude ratio.

The 2-state Potts model is equivalent to the Ising model which was
solved exactly~\cite{Onsager} (see Ref.~\cite{Ising-book} for
details). The susceptibility behavior was understood in the paper by Wu,
McCoy, Tracy and Barouch~\cite{MWTB}. It turns out that there exist
only integer corrections to scaling (for a recent and detailed
discussions we refer readers to Refs.~\cite{ONGP,CHPV,Delfino99}).
Values of the critical exponents and of some amplitude ratios are
presented in Table~\ref{tab2}.

\begin{table}
\caption{Exact values of critical exponents and ratios of critical
amplitudes for the Ising model (2-state Potts model).} \center
\begin{tabular}{|c|c|c|c|c|c|c|} \hline
$\nu$ & $\alpha$ & $\beta$ & $\gamma$ & $A_+/A_-$ & $\Gamma_+/\Gamma_L$ & $R_C^+=\Gamma_+A_+/B^2$ \\
\hline 1 & 0 & $1/8$ & $7/4$ & 1 & 37.69365... & 0.318569...
\\ \hline
\end{tabular}
\label{tab2}
\end{table}

The critical behavior of the susceptibility reads as
\begin{equation}
    \chi(\tau)=\Gamma_\pm\tmod^{-\gamma}
    {\cal X}_{corr}(\tmod^\Delta)+{\cal Y}_{bt}(\tau),
    \label{chi-gen} \end{equation}
\noindent where ${\cal X}_{corr}(\tmod^\Delta)$ is the
correction-to-scaling function and ${\cal Y}_{bt}(\tau)$ represents
an analytic expression (``background term'') which accounts for
non-singular contributions to susceptibility.

Set of values of the thermal critical exponents $x_{\epsilon_n}$ and
of the magnetic critical exponents $x_{\sigma_n}$ are known
analytically~\cite{Exponents,CFT}

\begin{equation}
x_{\epsilon_n}= \frac{n^2+ny}{2-y},\;\;\; x_{\sigma_n}=
\frac{\left(2n-1\right)^2-y^2}{4(2-y)} \label{set-of-exponents}
\end{equation}

\noindent in terms of the variable $y$ linked to the number of states
$q$ by $ \cos\frac{\pi y}{2}=\frac12\sqrt{q}$.

For the 3-state Potts model there is a finite number of correction
terms~\cite{CFT}, $x_{\epsilon_2}=14/5$, $x_{\epsilon_2}=6$, and
$x_{\sigma_2}=4/3$. The leading correction-to-scaling contribution
is $\Delta=-(2-x_{\epsilon_2})/(2-x_{\epsilon_1})=2/3$ and it was first
supported in numerical simulation~\cite{GRR85}.

\begin{figure}\centering
\caption{3-state Potts model. Difference of susceptibilities
$(\chi_L-\chi_T)\tmod^\gamma$ as function of $\tmod^{2/3}$. The almost
linear dependence supports the value 2/3 for the power of the
leading correction to scaling.}
\includegraphics[angle=270,width=.8\columnwidth]{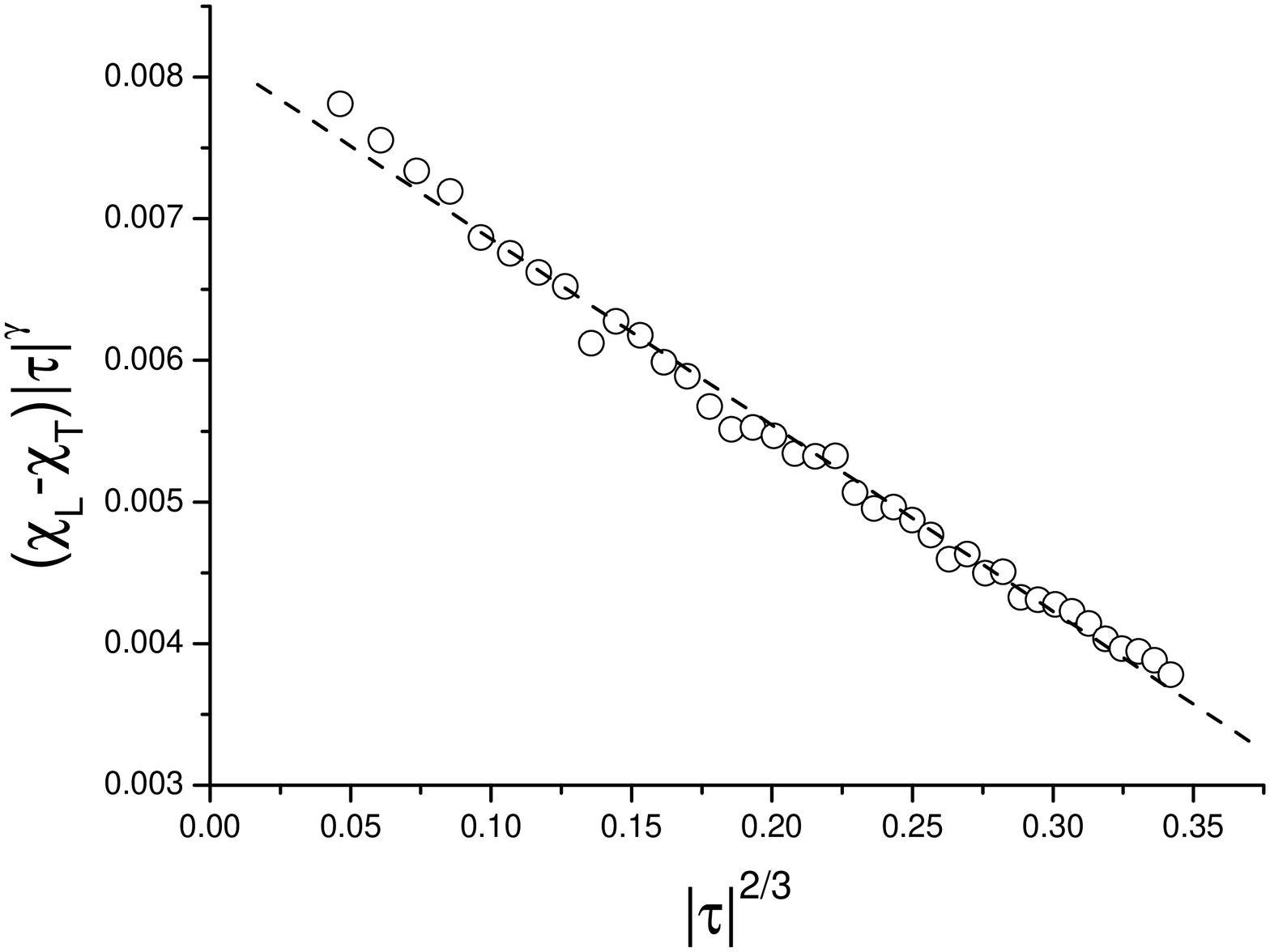}
\label{fig1}
\end{figure}

\begin{figure}\centering
\caption{3-state Potts model. Ratio of transverse to longitudinal
susceptibilities.}
\includegraphics[angle=270,width=.8\columnwidth]{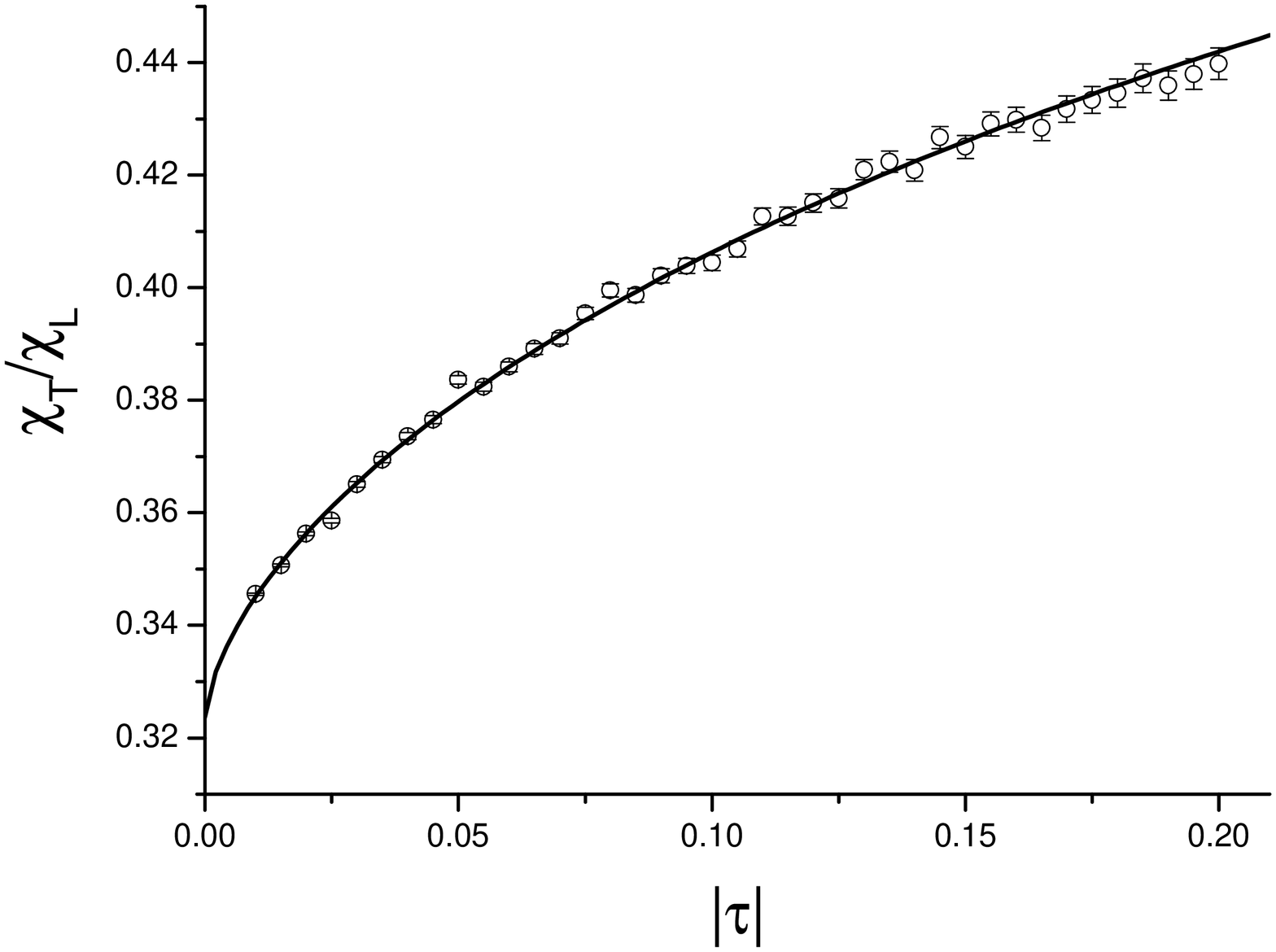}
\label{fig2}
\end{figure}

Clear evidence for these leading correction to the scaling behavior
may be seen in Figure~\ref{fig1}, where we plot the difference of the
longitudinal and transverse susceptibilities $(\chi_L-\chi_T)$ (note
the cancelation of background terms~(\ref{chi-gen}) in the
difference) multiplied by the leading behavior factor
$\tmod^\gamma$, as a function of $\tmod^{2/3}$. The ratio of the
susceptibilities $(\chi_T/\chi_L)$ is shown in Figure~\ref{fig2} and
may be used to estimate the universal ratio of associated amplitudes
$\Gamma_T/\Gamma_L$.

Analytical predictions for the amplitude ratios in Potts model for
$q=2,3$, and 4 were given in the
papers~\cite{DelfinoCardy98,DelfinoBarkemaCardy00}. The values are
shown in the Table~\ref{tab3} together with numerical estimations
from Monte Carlo (MC) and series expansions (SE)
analyses~\cite{SBB02,SBB08,EntGut}. The coincidence of the data is a
good indication for the validity of both two-kink
approximation~\cite{DelfinoCardy98,DelfinoBarkemaCardy00} to the
exact scattering theory~\cite{ChimZamolodchikov92} for 3-state Potts
model and of the analysis of MC data and series expansions.

\begin{table}
\caption{Exact values of critical exponents and estimates of the
ratios of critical amplitudes for the 3-state Potts model.} \center
\begin{tabular}{|c|c|c|c|l|l|l|l|} \hline
$\nu$ & $\alpha$ & $\beta$ & $\gamma$ & $\Gamma_+/\Gamma_L$ & $\Gamma_T/\Gamma_L$ & $R_C^+{=}\Gamma_+A_+/B^2$ & Remark\\
\hline 5/6 & 1/3 & $1/9$ & $13/9$ & - & - & - & Exact result \\
& & & & 13.848 & 0.327 & 0.1041 &
\cite{DelfinoCardy98,DelfinoBarkemaCardy00}\\
 & & & & 13.83(9) & 0.325(2) &  & \cite{EntGut} - SE \\
& & & & 13.83(9) & 0.3272(7) & 0.1044(8) & \cite{SBB08} - SE \\
& & & & 13.86(12) & 0.322(3) & 0.1049(29) & \cite{SBB08} - MC
\\ \hline
\end{tabular}
\label{tab3}
\end{table}

 The analysis of the 4-state Potts model is much more complicated because
in addition to the corrections to scaling there are confluent
logarithmic corrections~\cite{NS80,SalasSokal97}. The result of the
analysis of MC data and SE~\cite{CTV-SW,SBB-EPL,EntGut} is shown in
the Table~\ref{tab4} together with the analytical
estimates~\cite{DelfinoCardy98,DelfinoBarkemaCardy00}.

Conformal field theory predicts the set of renormalization group
(RG) exponents $y_{\epsilon_n}{=}2{-}x_{\epsilon_n}{=}$3/2, 0, -5/2,
... . The leading correction-to-scaling exponent $y_{\epsilon_2}$
vanishes and gives rise to a logarithmic behavior~\cite{NS80}. In
our recent publication~\cite{SBB-EPL,SBB-4} we revised
the renormalization group equations and included in our analysis the known
form of the logarithmic corrections and of the next-to-leading corrections,
taking into account the width of the temperature region window examined.
The set of magnetic exponents for the Potts model $x_{\sigma_n}${=}1/8, 9/8,
25/8, ... translates into the magnetization exponent $\beta{=}1/12$
and the leading correction-to-scaling exponent $\Delta_\sigma{=}2/3$.
Finally, the following behavior of the susceptibility is assumed
\begin{equation}
    \chi_+(\tau) =
    \Gamma_+\tau^{-7/6}{\cal G}^{3/4}(-\ln\tau)
    (1+a_{+}\tau^{2/3}+b_+\tau)
    +D_+,
    \label{chi-4-sing}
\end{equation}
\noindent where the function $\cal G$ contains a universal correction
function $\cal E$~\cite{SalasSokal97,SBB-EPL,SBB-4} and the leading
nonuniversal correction function $\cal F$ \bey
    {\cal G}(-\ln|\tau|)&=&(-\ln|\tau|)\times{\cal E}(-\ln|\tau|)\times {\cal F}(-\ln|\tau|),
    \label{Eq-Dfn_H}\\
    {\cal E}(-\ln|\tau|)&=&
    \left(1+\frac 34\frac{\ln(-\ln|\tau|)}{-\ln|\tau|}\right)
        \nonumber\\
    &&
    \ \times\left(1-\frac 34\frac{\ln(-\ln|\tau|)}{-\ln|\tau|}\right)^{-1}
    \left(1+\frac 34\frac{1}{(-\ln|\tau|)}\right),
        \label{Eq-Dfn_G}\\
    {\cal F}(-\ln|\tau|)&\simeq&
    \left(1+\frac{C_1}{-\ln|\tau|}
    +\frac{C_2\ln(-\ln|\tau|)}{(-\ln|\tau|)^2}\right)^{-1}.\label{eq-f59}
\eey

We fit our data to estimate the amplitude $\Gamma_+$, the coefficient of
the leading correction to scaling $a_+$ in Eq.~(\ref{chi-4-sing}),
and coefficients $C_1$ and $C_2$ in Eq.~(\ref{eq-f59}).

It is obvious that the logarithmic corrections (the whole function
$\cal G(-\ln\tmod)$) cancels in simple ratios, like $A_+/A_-$,
$\Gamma_+/\Gamma_L$, $\Gamma_T/\Gamma_L$, etc. This has been
demonstrated analytically for the effective ratio $A_+/A_-$ (see
Eq.~\ref{d-e-linear}). We note also that the RG analysis
predicts~\cite{SBB-EPL,SBB-4} powers of logarithmic corrections to
specific heat $\alpha'{=}{-}1$, susceptibility $\gamma'{=}3/4$, and
magnetization $\beta'{=}{-}1/8$  such  that they cancel in all
universal ratios. For example, the universal amplitude ratio $R_C^-$
may be calculated as the limit of the ratio of functions

\begin{equation}
R_{C}^-=\lim_{\tau\rightarrow 0}
\tau\frac{(E_-(\tmod)-E_0)\chi_-(\tmod)}{M(\tmod)^2}
\alpha(\alpha-1)\beta_c \label{Rcp}
\end{equation}
\noindent where $E_0=E(0)=\sqrt{2}$. One can check that in the ratio
(\ref{Rcp}) not only powers of $\tmod$ cancel but also powers of $\cal E$.
 In the ratio, the magnetization $M$ and the energy
difference $E_-(\tmod)-E_0$ have only singular contributions and the
only systematic deviation may come from the background correction to
susceptibility $\chi_-(\tmod)$. It was shown in~\cite{SBB08} that
the contribution from this background correction is negligible in
the critical region window, and the estimator~(\ref{Rcp}) tends to
the value $0.0055(1)$ as $\tau\rightarrow 0$.

The Baxter-Wu~\cite{Baxter-Wu} model is defined on a triangular lattice,
with spins $\sigma_i$ located at vertices. The three spins forming a
triangular face are coupled with a strength $J$, and the Hamiltonian
reads

\begin{equation}
{\cal H} =-J\sum_{\rm faces} \sigma_i\sigma_j\sigma_k,
\label{ham-bw}
\end{equation}

\noindent where the summation extends over all triangular faces of
the lattice, both pointing up and down. The ground state is four-fold
degenerate and the critical exponents are found to be the same
as for the 4-state Potts model.

The exact behavior of the magnetization, energy, and specific heat
are known analytically~\cite{Baxter-Wu,Baxter-Book,Joyce}. An analysis
of Monte Carlo data was performed by two of us~\cite{LW} and preliminary
estimates shows that the values of the susceptibility amplitude-ratio
$\Gamma_+/\Gamma_L\approx 6.9$ and of the ratio $R_C^-\approx 0.005$
 are very near
 to those  obtained from our analysis of MC and SE data for the
4-state Potts model (see Table~\ref{tab4}). We have to note that
logarithmic corrections to scaling are absent in the critical
behavior of Baxter-Wu model and this gives us more confidence in our
analysis.

Delfino and Grinza~\cite{DG04} use the same analytical approach as
in~\cite{DelfinoCardy98} to study the Ashkin-Teller model which also
belongs to the 4-state Potts model universality class with some
particular choice of coupling constants. This leads to the
estimatation $\Gamma_+/\Gamma_L\approx 4.02$ and
$\Gamma_T/\Gamma_L\approx 0.129$. This result is also very near to
those for 4-state Potts model (see second entry in
Table~\ref{tab4}.)

A possible explanation of the deviation of our results from the
analytical predictions may be explained as follows: the two-kink
approximation is exact for the 2-state Potts model, it gives good accuracy
for the 3-state Potts model, but it may be insufficient to produce accurate
values for 4-state Potts model. Further analyses have to be done to
resolve the contradiction  among these results.

\begin{table}
\caption{Exact values of critical exponents and estimations of the
ratios of critical amplitudes for the 4-state Potts model.} \center
\begin{tabular}{|c|c|c|c|l|l|l|l|} \hline
$\nu$ & $\alpha$ & $\beta$ & $\gamma$ & $\Gamma_+/\Gamma_L$ & $\Gamma_T/\Gamma_L$ & $R_C^-{=}\Gamma_-A_-/B^2$ & Remark\\
\hline 2/3 & 2/3 & $1/12$ & $7/6$ & - & - & - & Exact result \\
& & & & 4.013 & 0.129 & 0.00508 &
\cite{DelfinoCardy98,DelfinoBarkemaCardy00}\\
& & & & 3.5(4) & 0.11(4) &  & \cite{EntGut} - SE \\
& & & & 3.14(70) &  & 0.0068(9) & \cite{CTV-SW} - MC \\
& & & & 6.93(6) & 0.1674(30) & 0.00512(13) & \cite{SBB-EPL,SBB-4} - MC \\
& & & & 6.30(1) & 0.1511(24) & 0.00531(5) & \cite{SBB-EPL,SBB-4} -
SE
\\ \hline
\end{tabular}
\label{tab4}
\end{table}

BB and WJ acknowledge partial support within the Graduate School
``Statistical Physics of Complex Systems'' of DFH-UFA under Contract
No.\ CDFA-02-07. Financial support within a common research program
between the Landau Institute and the Ecole Normale Sup\'erieure de
Paris, Paris Sud University is also gratefully acknowledged.



\end{document}